\documentclass[final]{IEEEtran}
\IEEEoverridecommandlockouts
\usepackage{cite}
\usepackage{amsmath,amssymb,amsfonts}
\usepackage{algorithmic}
\usepackage{graphicx}
\usepackage{url}
\usepackage{textcomp}
\usepackage{xcolor}
\usepackage{float}
\def\BibTeX{{\rm B\kern-.05em{\sc i\kern-.025em b}\kern-.08em
    T\kern-.1667em\lower.7ex\hbox{E}\kern-.125emX}}

\floatstyle{plain}
\newfloat{copyrightbox}{!hb}{loc}
\floatname{copyrightbox}{Copyright Notice}

\newcommand\copyrighttext{%
\footnotesize 
\textit{This is a preprint accepted at the 22nd International Conference on Machine Learning and Applications (ICMLA), 2023. \textcopyright IEEE.}
}

\begin{document}

\title{Unveiling the Potential of Deep Learning Models for Solar Flare Prediction in Near-Limb Regions}
\author{\IEEEauthorblockN{Chetraj Pandey, Rafal A. Angryk, Berkay Aydin}
\IEEEauthorblockA{\textit{Department of Computer Science, Georgia State University, Atlanta, GA, USA} \\
\textit{\{cpandey1, rangryk, baydin2\}@gsu.edu}\\
} }

\maketitle

\begin{abstract}
% In recent years, the development of complex models for data-driven solar flare prediction has been accelerated by advancements in machine learning and deep learning utilizing a variety of approaches and data products, while most studies only address and assess the models' efficacy in central locations (within $\pm$70$^{\circ}$ in longitude of the solar disk).
This study aims to evaluate the performance of deep learning models in predicting $\geq$M-class solar flares with a prediction window of 24 hours, using hourly sampled full-disk line-of-sight (LoS) magnetogram images, particularly focusing on the often overlooked flare events corresponding to the near-limb regions (beyond $\pm$70$^{\circ}$ of the solar disk). We trained three well-known deep learning architectures--AlexNet, VGG16, and ResNet34 using transfer learning and compared and evaluated the overall performance of our models using true skill statistics (TSS) and Heidke skill score (HSS) and computed recall scores to understand the prediction sensitivity in central and near-limb regions for both X- and M-class flares. The following points summarize the key findings of our study: (1) The highest overall performance was observed with the AlexNet-based model, which achieved an average TSS$\sim$0.53 and HSS$\sim$0.37; (2) Further, a  spatial analysis of recall scores disclosed that for the near-limb events, the VGG16- and ResNet34-based models exhibited superior prediction sensitivity. The best results, however, were seen with the ResNet34-based model for the near-limb flares, where the average recall was approximately 0.59 (the recall for X- and M-class was 0.81 and 0.56 respectively) and (3) Our research findings demonstrate that our models are capable of discerning complex spatial patterns from full-disk magnetograms and exhibit skill in predicting solar flares, even in the vicinity of near-limb regions. This ability holds substantial importance for operational flare forecasting systems.
\end{abstract}

\begin{copyrightbox}
\centering
  \fbox{\parbox{\dimexpr\linewidth-5\fboxsep-3\fboxrule\relax}{\copyrighttext}}
\end{copyrightbox}
\section{Introduction}\label{sec:int}
Solar flares are temporary occurrences on the Sun, considered to be the central phenomena in space weather forecasting, manifested as the sudden large eruption of electromagnetic radiation on the outermost atmosphere of the Sun. They are observed by the X-ray sensors on Geostationary Operational Environmental Satellite (GOES) and classified according to their peak X-ray flux level (measured in watt per square meter) into the following five categories by the National Oceanic and Atmospheric Administration (NOAA): X $(\geq10^{-4}Wm^{-2})$, M $(\geq10^{-5}$ and $< 10^{-4}Wm^{-2})$, C $(\geq10^{-6}$ and $<10^{-5}Wm^{-2})$, B $(\geq10^{-7}$ and $<10^{-6}Wm^{-2})$, and A $(\geq10^{-8}$ and $<10^{-7}Wm^{-2})$ \cite{spaceweather}; areas on the Sun with peak X-ray flux less than  $<10^{-8}Wm^{-2}$ (less than A-class flares) are regarded as flare-quiet (FQ) regions. Large flares (M- and X-class) are rare events and significantly more powerful than other flare classes, with the capability to disrupt several infrastructures in space (e.g., satellite operations) and on Earth (e.g., the electricity power grid and avionics). Hence, a precise and reliable system for solar flare prediction is essential.

\begin{figure}[tbh!]
\centering
\includegraphics[width=0.9\linewidth ]{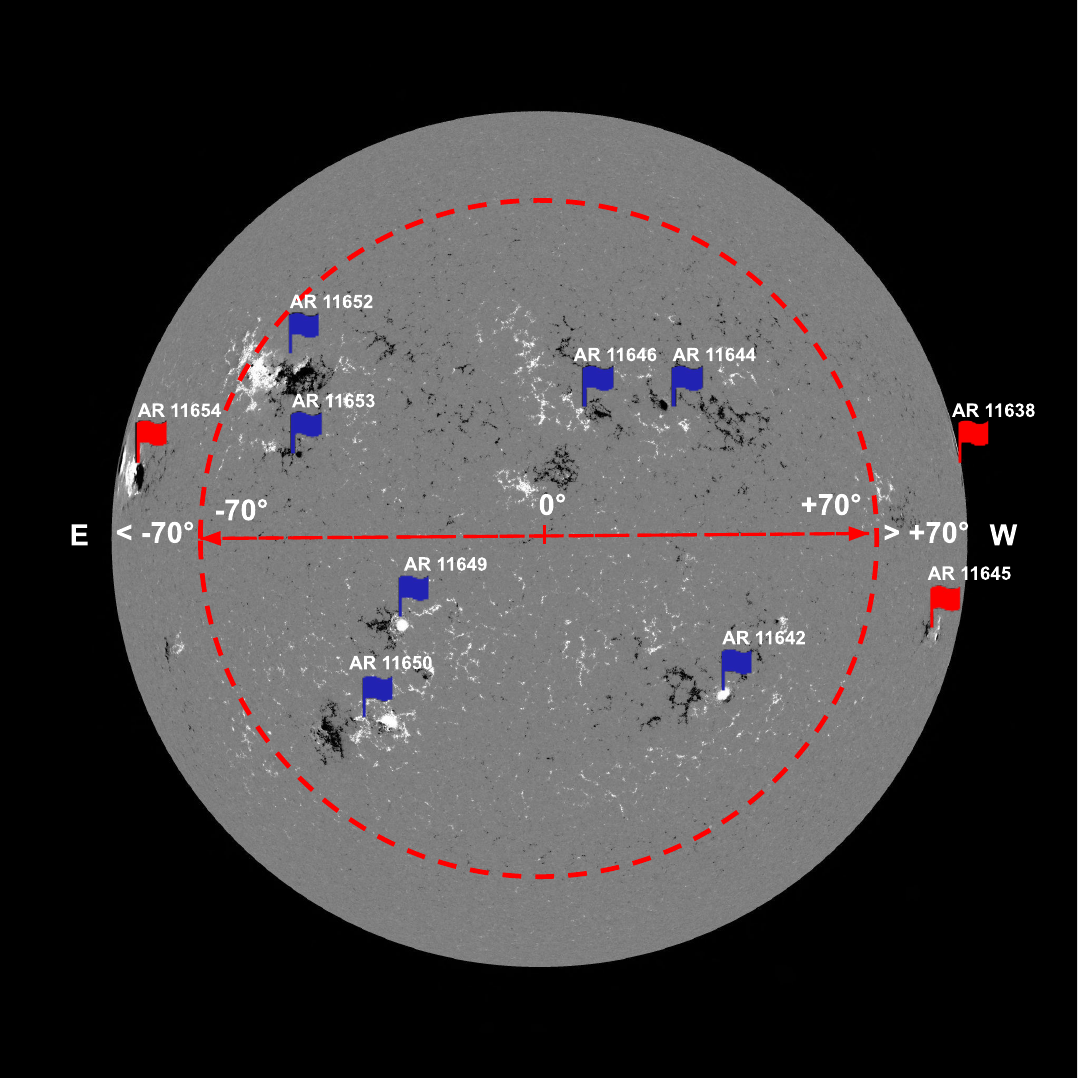}
% \caption[]{An annotated full-disk LoS magnetogram image observed on 2011-12-28 at 12:00:00 UTC as an example, showing the central location (within $\pm$70$^{\circ}$) and near-limb (beyond $\pm$70$^{\circ}$ to $\pm$90$^{\circ}$) region. Note that the directions East (E) and West (W) are reversed in solar coordinates.}
\caption[]{An annotated full-disk line-of-sight magnetogram observed on 2013-01-09 at 00:00:00 UTC as an example, showing the approximate central location (within $\pm$70$^{\circ}$) and near-limb (beyond $\pm$70$^{\circ}$ to $\pm$90$^{\circ}$) region with all the NOAA active regions (ARs) present at the noted timestamp. ARs in central and near-limb regions are indicated by blue and red flags respectively. Note that the directions East (E) and West (W) are reversed in solar coordinates.}
\label{fig:mag}
\vspace{-5pt}
\end{figure}

Active regions (ARs) on the Sun are areas with localized magnetic disturbances, visually indicated by scattered flags in the full-disk magnetogram image shown in Fig.~\ref{fig:mag}. In most operational flare forecasting systems, these ARs are used as regions of interest because they are considered to be the main initiators of space weather events. In AR-based flare prediction, the underlying models issue predictions for each AR individually. To issue a full-disk forecast with an AR-based model, the output flare probabilities for each active region are usually aggregated using a heuristic function as mentioned in \cite{Pandey2022f}. This heuristic function assumes conditional independence among ARs and equal contributions from all ARs to the aggregated full-disk forecast. However, this uniform weighting scheme may not accurately represent the actual impact of each AR on the probability of full-disk flare prediction \cite{pandey2022exploring}. % To issue a full-disk forecast with an AR-based model, the output flare probabilities for each active region are usually aggregated using a heuristic function as mentioned in \cite{Pandey2022f}. The heuristic function used to aggregate the final forecast operates under the assumption of conditional independence among active regions and that all active regions contribute equally to the aggregate forecast. This uniform weighting scheme may not accurately reflect the true influence of each active region on full-disk flare prediction probability. It's important to highlight that the weights of these active regions are generally unknown; there are no established methods to accurately determine them, nor are there any prior assumptions that guide the assignment of these weights. 
Furthermore, the magnetic field measurements, which are the dominant feature employed by the AR-based forecasting techniques, are susceptible to severe projection effects \cite{Hoeksema2014} as ARs get closer to limbs (to such an extent that magnetic field readings become distorted beyond $\pm$60$^{\circ}$ of the solar-disk \cite{Falconer2016}). Therefore, the aggregated full-disk flare probability is in fact, limited to ARs in central locations. This further underscores the inherent challenges in issuing a full-disk flare forecast using an AR-based model. As studies in AR-based model for flare prediction include ARs located within $\pm$30$^{\circ}$ \cite{Huang2018} to $\pm$70$^{\circ}$ \cite{Ji2020}, in the context of this paper, this upper limit ($\pm$70$^{\circ}$) is used as a boundary for central location (within $\pm$70$^{\circ}$)  and near-limb regions (beyond $\pm$70$^{\circ}$) as shown in Fig.~\ref{fig:mag}. Full-disk models, on the other hand, typically use compressed images derived from the original full-depth magnetogram rasters representing the entire solar disk. While projection effects persist in these magnetograms, which contain actual magnetic field readings, the compressed images retain only the shape-based spatial patterns of AR such as size, directionality, sunspot borders, and inversion lines in grayscale. Thus, by incorporating the entire full-disk magnetogram, this approach enables the prediction of solar flares in the Sun's near-limb areas as well, which are often overlooked by AR-based methods. 

% and enhancing operational systems.To effectively analyze this information, convolutional neural networks (CNNs), are proven to be adept at capturing spatial patterns and relationships, the  The CNN models can automatically extract relevant spatial features and discern the intricate structures and configurations associated with active regions prone to solar flares. This enables the CNNs to recognize and learn the specific spatial characteristics that indicate an elevated probability of solar flare occurrence. In essence, this approach harnesses the power of deep learning for spatial analytics to interpret the underlying patterns and variations within full-disk line-of-sight (LoS) magnetograms, thereby enabling robust prediction of solar flares. While projection effects persist in these images, it remains to be proven whether full-disk models are capable of predicting flares from areas close to the near-limb. Thus, we provide quantitative evidence favoring a full-disk model and show that it is essential to supplement AR-based models, enabling the prediction of flares in the Sun's near-limb areas and enhancing operational systems.

To the best of our knowledge, flare prediction employs four major approaches: (i) empirical human prediction (e.g., \cite{Crown2012}), (ii) statistical prediction (e.g., \cite{Lee2012}), (iii) physics-based numerical simulations (e.g., \cite{Kusano2020}), and (iv) machine learning and deep learning approaches (e.g., \cite{Li2020, Ji2020, Whitman2022, pandeyds2023, pandeydsaa2023, Hong2023}). The use of machine learning in extracting forecast patterns from the Sun has been an active area of research since the early 1990s \cite{Aso1994}.  Since then, there has been a significant advancement in machine learning and deep learning techniques, leading to a surge of interest in applying these methods to build more precise flare forecasting models as they can automatically extract relevant spatial features discerning the intricate structures associated with ARs prone to eventual solar flares.  
% For example, in \cite{Nishizuka2018}, a multi-layer perceptron-based model was presented. This model used manually selected 79 solar features to predict $\geq$C- and $\geq$M-class flares.

In \cite{Huang2018}, a convolutional neural network (CNN) model was trained using solar Active Region (AR) patches extracted from line-of-sight (LoS) magnetograms within $\pm$30$^{\circ}$ of the central meridian to predict $\geq$C-, $\geq$M-, and $\geq$X-class flares. Similarly, \cite{Li2020} developed a CNN-based model that issued binary class predictions for $\geq$C- and $\geq$M-class flares within 24 hours using Space-Weather Helioseismic and Magnetic Imager Active Region Patches (SHARP) data \cite{Bobra2014}. The SHARP data was extracted from solar magnetograms using AR patches located within $\pm45^{\circ}$ of the central meridian. Notably, both models \cite{Huang2018, Li2020} had limited operational capability, as they were restricted to small portions of the observable disk in central locations ($\pm30^{\circ}$ and $\pm45^{\circ}$). More recently, we presented full-disk models trained with limited data in \cite{Pandey2021, Pandey2022}. However, these were preliminary studies that did not provide insights into the model's capability for near-limb events. 

Moreover, our prior work on explainable full-disk deep learning models \cite{pandeyecml2023, pandeyds2023} shows that the features learned by these models are linked to relevant ARs on full-disk magnetograms which underscore their possible implications as a complementary approach. In this study, we present a more comprehensive view of deep learning-based full-disk models for predicting $\geq$M-class solar flares in binary mode. We evaluate and compare the performance of three widely used pre-trained CNNs – AlexNet \cite{alex}, VGG16 \cite{vgg}, and ResNet34 \cite{resnet} – by utilizing hourly sampled instances of full-disk LoS magnetogram images covering solar cycle 24. The focus of this work is to study whether our models can be relied upon for critical applications, particularly for near-limb forecasting with quantitative spatial analytics. The novel contributions of this paper are as follows: (i) We show an improved overall performance of a full-disk solar flare prediction model by building and comparing three CNN architectures on full-disk magnetogram images, (ii) We provide an extended spatial analysis of the predictive capability of full-disk models on near-limb and central locations, and (iii) We provide results that underscore the role of full-disk models in the prediction of solar flares in near-limb regions of the Sun.

The remainder of the paper is structured as follows. In Sec.~\ref{sec:data}, we detail the process of data preparation with consequent data distribution. In Sec.~\ref{sec:model}, we describe all three model architectures explored in this work. Sec.~\ref{sec:expt} presents the experimental design and model evaluation with skill scores and provides models' prediction sensitivity in central and near-limb regions. Finally, in Sec.~\ref{sec:conc}, we summarize our findings and suggest avenues for future research.

\section{Data} \label{sec:data}
\begin{figure}[tbh!]
\centering
\vspace{-10pt}
\includegraphics[width=0.95\linewidth ]{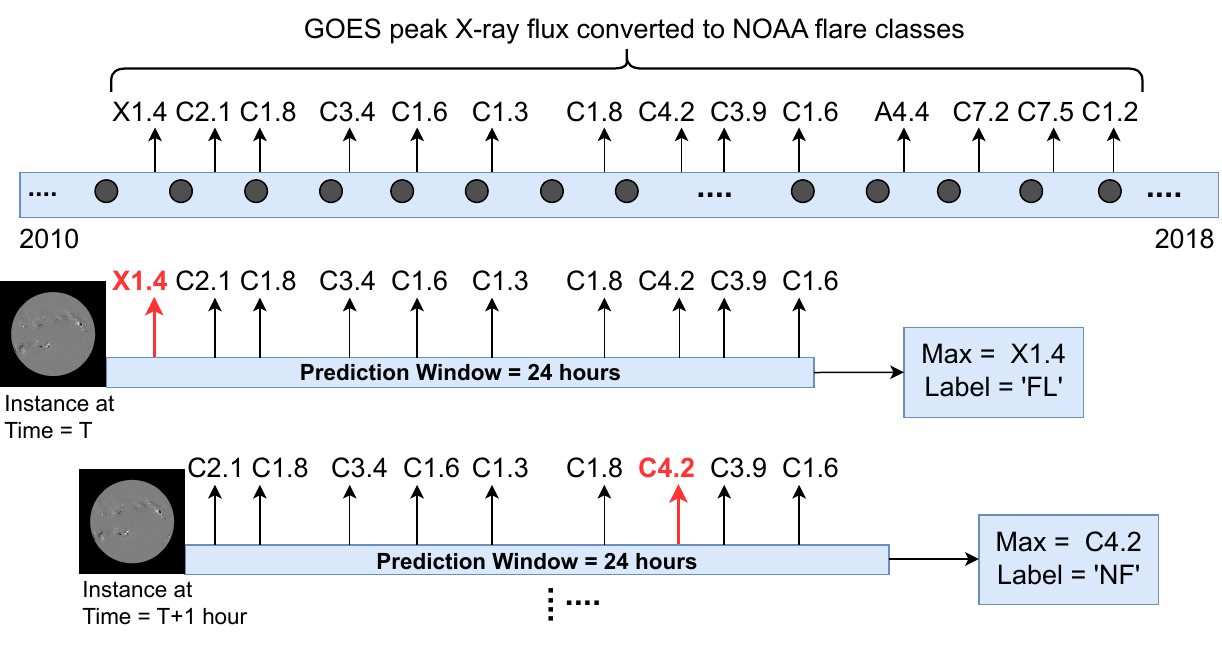}
\caption[]{An illustration of data labeling process for hourly observations of full-disk LoS magnetogram images with a prediction window of 24 hours. Here, `FL'  and `NF' indicate `Flare' and `No Flare' classes. The gray-filled circles indicate hourly spaced timestamps for magnetogram instances.}
\label{fig:timeline}
\vspace{-10pt}
\end{figure}

\begin{figure}[tbh!]
\centering
\includegraphics[width=0.95\linewidth ]{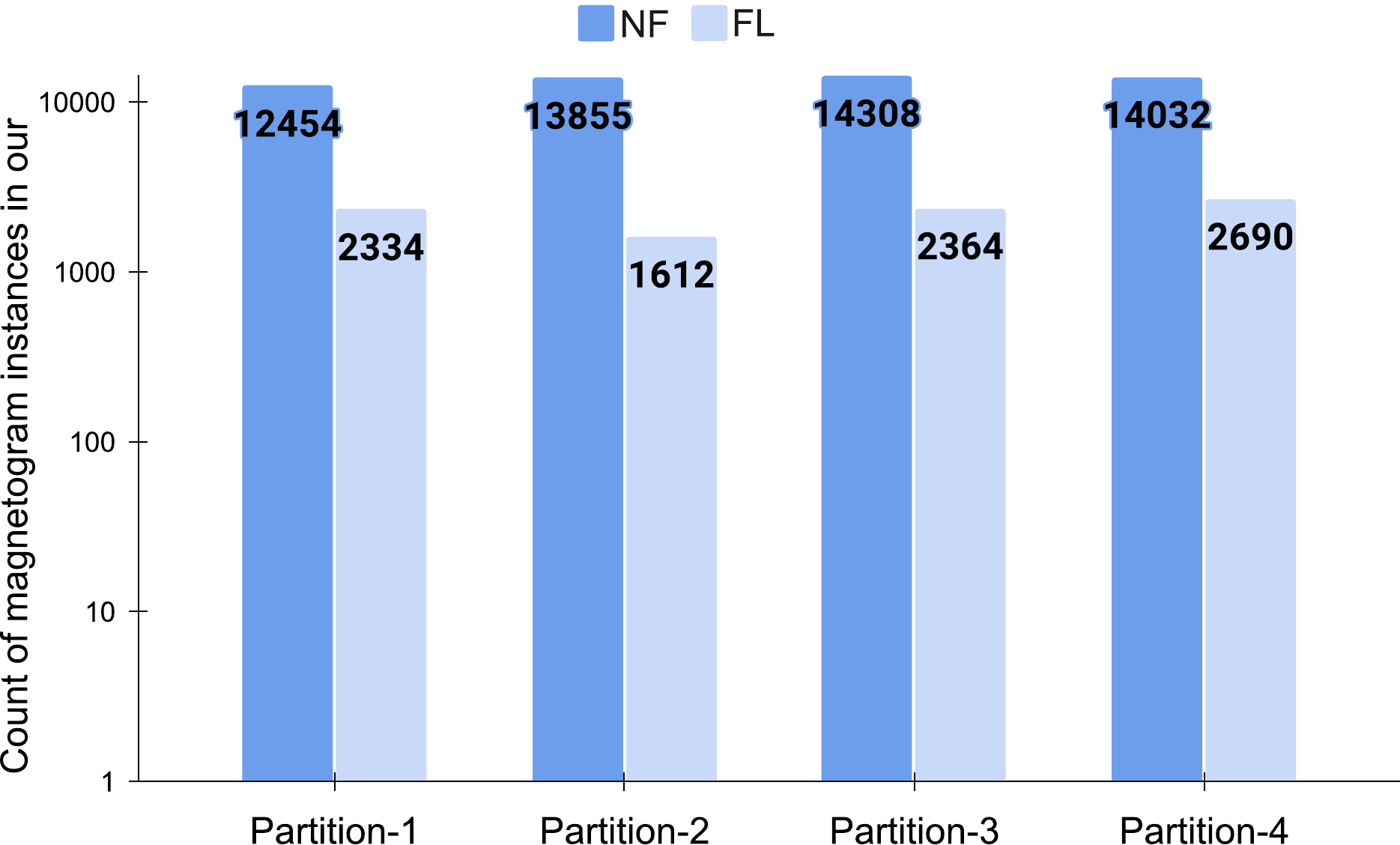}
\caption[]{Data distribution of four tri-monthly partitions for predicting $\geq$M1.0-class flares. Note that the length of the bars are in logarithmic scale.}
\label{fig:data}
\vspace{-5pt}
\end{figure}
% \begin{table}[t!]
% \setlength{\tabcolsep}{4pt}
% \renewcommand{\arraystretch}{1.75}
% \caption{The total number of hourly sampled magnetogram images per flare class distributed into four tri-monthly partitions.}
% \begin{center}
% \begin{tabular}{c c c c c c}
% \hline
% % & & \textbf{Loss}& \textbf{Learning} & \textbf{Batch}& \\
% Binary Class & Partition-1 & Partition-2 & Partition-3 & Partition-4 & Total\\
% \hline
% NF ($<$M1.0) & 12,454 & 13,855 & 14,308 & 14,032 & 54,649 \\
% FL ($\geq$M1.0) & 2,334 & 1,612 & 2,364 & 2,690 & 9,000 \\
% \hline
% FL:NF & $\sim$1:5 & $\sim$1:9 & $\sim$1:6 & $\sim$1:5 & $\sim$1:6 \\
% \hline
% \end{tabular}
% \label{table:datatable}
% \end{center}
% \vspace{-17pt}
% \end{table}

% \begin{figure*}[ht!]
% \begin{tabular}{c c}
%     \includegraphics[width=0.45\linewidth]{figures/data.pdf} & 
% \includegraphics[width=0.45\linewidth]{figures/partition.pdf} \\
%     (a) & (b)
% \end{tabular}
% \caption{(a) The total number of hourly sampled magnetograms images per flare classes. (b) Label distribution into four tri-monthly partitions for predicting $\geq$M1.0-class flares. Note that the length of the bar in both of the figures is in logarithmic scale.}
% \label{fig:data}

% \end{figure*}

We use compressed images of full-disk LoS solar magnetograms obtained from the Helioseismic and Magnetic Imager (HMI) \cite{Schou2011} instrument onboard Solar Dynamics Observatory (SDO) \cite{Pesnell2011}, publicly available from Helioviewer\footnote{Helioviewer : \url{https://api.helioviewer.org/}}. We collected a total of 63,649 magnetogram images by sampling them every hour of the day, starting at 00:00 and ending at 23:00, from December 2010 to December 2018. These images are resized to 512$\times$512 for computational efficiency and labeled using a 24-hour prediction window based on the maximum peak X-ray flux (converted to NOAA flare classes), as illustrated in Fig.~\ref{fig:timeline}. To elaborate, if the maximum X-ray intensity of a flare was weaker than M, we labeled the observation as "No Flare" (NF: $<$M), and if it was $\geq$M, we labeled it as "Flare" (FL: $\geq$M). This resulted in 54,649 instances for the NF class and 9,000 instances (8,120 instances of M-class and 880 instances of X-class flares) for the FL class. Due to scarce M- and X-class flares, the overall class imbalance in our dataset is $\sim$1:6 (FL:NF). Finally, we created a non-chronological split of our data into four temporally non-overlapping tri-monthly partitions introduced in \cite{Pandey2021} for our 4-fold cross-validation experiments, where three of the partitions are used as a training set, and one partition is used as a test set. The detailed class-wise data distribution is shown in Fig.~\ref{fig:data} 

\section{Models} \label{sec:model}
In this work, we use three CNN architectures: AlexNet \cite{alex}, VGG16 \cite{vgg}, and ResNet34 \cite{resnet}. We used AlexNet \cite{alex} due to its inherent architectural simplicity, consisting of 5 convolutional layers, 3 max pool layers, 1 adaptive average pool layer, and three fully connected layers. Moreover, our study included VGG16 \cite{vgg}, a relatively more complex model, to evaluate the hypothesis that an increase in the number of layers might engender enhanced performance. This model augments the foundational structure of AlexNet by integrating additional convolutional layers, all employing uniform 3x3 convolutional kernels. The VGG16 architecture consists 13 convolutional layers, 5 max pool layers, 1 adaptive average pool layer, and 3 fully connected layers. Lastly, we included ResNet34 \cite{resnet}, a CNN model that extends the complexity of the VGG16 design by facilitating the training of deeper networks with fewer parameters. Diverging from the approach of AlexNet and VGG16, ResNet34 integrates residual connections from each layer into subsequent connected layers. The architecture of ResNet34 consists of 33 convolutional layers, including a 7x7 kernel for the initial layer and 3x3 kernels for the remaining layers, along with one max pool layer, one adaptive average pool layer, and one fully connected layer. 
% \begin{figure}[htbp!]
% \centering
% \includegraphics[width=0.98\linewidth ]{figures/icmla_model.pdf}
% \caption[]{An overview of three deep learning architectures we use (a) AlexNet-, (b) VGG16-, (c) ResNet34-based models.}
% \label{fig:arch}
% \end{figure}

The primary reason behind our choice of these three architectures was to analyze and evaluate the influence of varying architectural designs and increasing layer depths on performance. Additionally, we factored the simplicity of the architectures into our selection process, in light of the relatively modest scale of our dataset for deep learning models. These pre-trained models require a 3-channel image for input, however, our data comprises compressed solar magnetogram images, which are grayscale. To reconcile this, we incorporated an additional convolutional layer at the onset of the network architecture, which accepts a 1-channel input. This layer employs a 3$\times$3 kernel with a size-1 stride, padding, and dilation, and consequently generates a 3-channel feature map. We initialize the added convolutional layer in all three models using Kaiming initialization \cite{kaiming}, while all other layers are initialized with pre-trained weights.
% Additionally, with the aim of optimally utilizing the pre-trained weights—-irrespective of the architectural specifics of these models, which anticipate 3-channel input of varying dimensions—-we used an adaptive average pooling layer within each model. This layer is positioned after the completion of feature extraction via the convolutional layer and immediately preceding the fully-connected layer. This placement facilitates the alignment of dimensions with our image input size, which is 512$\times$512.

\section{Experimental Evaluation} \label{sec:expt}
% In this section, we provide a comprehensive overview of our experimental setup, outlining the settings for data augmentation techniques employed for data balancing, and the hyperparameter configurations utilized to train our models. Moreover, we present the obtained results and share our observational remarks derived from the experiments, with a specific emphasis on the crucial aspect of flare spatial locations, specifically near-limb flares. These near-limb flares are often overlooked, and our analysis sheds light on the predictive capabilities of our models in operational systems.

\subsection{Experimental Settings}
% \begin{table*}[tb]
% \setlength{\tabcolsep}{8pt}
% \renewcommand{\arraystretch}{1.5}
% \caption{Parameter settings for all of our models: AlexNet, VGG16, and ResNet34.}
% \begin{center}
% \begin{tabular}{c c c c c c c c}
% \hline
% % & & \textbf{Loss}& \textbf{Learning} & \textbf{Batch}& \\
%  &  & \textbf{Loss}& \textbf{Initial} & \textbf{Max. Learning} & \textbf{Batch}& \textbf{Weight}& \\
% \textbf{Models} &\textbf{Optimizer}  & \textbf{Function}& \textbf{Learning Rate} & \textbf{Rate}& \textbf{Size} & \textbf{Decay}& \textbf{Epochs}\\
% \hline
% AlexNet & SGD& NLL & $1e-5$& $1e-4$& 64 & $1e-4$ & 50\\
% VGG16 & SGD& NLL & $1e-5$& $1e-5$& 64 & $1e-4$ & 50\\
% ResNet34 & SGD& NLL & $1e-5$&$1e-5$& 64 & $1e-3$ & 50\\
% \hline
% \end{tabular}
% \label{table:param}
% \end{center}
% \vspace{-10pt}
% \end{table*}
We trained our full-disk flare prediction models using Stochastic Gradient Descent (SGD) as the optimizer and Negative Log-Likelihood (NLL) as the objective function. We initialized each of the models with their corresponding pre-trained weights, then further trained it for 50 epochs while employing a dynamic learning rate scheduling strategy, OneCycleLR with cosine annealing \cite{onecyclelr}. The initial learning rate (LR) employed for all three models was $1e-5$, with a maximum LR set to $1e-5$ for VGG16 and ResNet34, and $1e-4$ for the AlexNet model. This scheduler adjusts the learning rate in a cyclical pattern, gradually increasing it to help the model quickly converge and then decreasing it to fine-tune the performance. The steps per epoch were set to the number of batches in training data, and the batch size was 64. Using the OneCycleLR scheduler simplifies and optimizes the models' learning rate schedule for hyperparameter tuning.

As mentioned earlier in Sec.~\ref{sec:data}, our dataset has an inherent class imbalance issue. This imbalance can significantly influence the performance of the models, potentially leading to less precise and reliable predictions for the minority class. To address this, we employed data augmentation and adjusted class weights in the loss function only on the training set. Specifically, we applied three augmentation techniques: vertical flipping, horizontal flipping, and rotations between +5$^{\circ}$ and -5$^{\circ}$ to both classes. For each instance in the minority class (FL), we applied all three augmentations, quadrupling the total number of instances for the entire FL-class. For each instance in NF-class, we randomly selected one of the three aforementioned augmentation techniques, doubling the total instances for this class. The goal of augmenting the NF-class instances was to ensure that the NF-class, though not uniformly augmented, retained a diversity in its data akin to the FL-class and expanded the overall dataset. Post augmentation, we adjusted the class weights to be inversely proportional to class frequencies, thereby penalizing misclassifications of the minority class. Finally, we evaluated our models using a 4-fold cross-validation approach on tri-monthly partitions.

We evaluate the overall performance of our models using two widely-used forecast skills scores: true skill statistic (TSS, in Eq.~\ref{eq:TSS}) and Heidke skill score (HSS, in Eq.~\ref{eq:HSS}), derived from the elements of confusion matrix: True Positives (TP), True Negatives (TN), False Positives (FP), and False Negatives (FN). In the context of this paper, the FL class is considered as the positive outcome, while NF is negative. Lastly, we report the subclass and overall recall (shown in Eq.~\ref{eq:rec}) for flaring instances (M- and X-class) to assess the prediction sensitivity of our models in central and near-limb regions. 

\begin{equation}\label{eq:TSS}
    TSS = \frac{TP}{TP+FN} - \frac{FP}{FP+TN}
\end{equation}

\begin{equation}\label{eq:HSS}
    HSS = 2\times \frac{TP \times TN - FN \times FP}{((P \times (FN + TN) + (TP + FP) \times N))},
\end{equation}
\begin{center}
\vspace{2pt}
    where $N = TN + FP$ and $P = TP + FN$.
\end{center} 

\begin{equation}\label{eq:rec}
    Recall = \frac{TP}{TP+FN}
    \vspace{2pt}
\end{equation}

TSS and HSS values range from -1 to 1, where 1 indicates all correct predictions, -1 represents all incorrect predictions, and 0 represents no skill. In contrast to TSS, HSS is an imbalance-aware metric, and it is common practice to use HSS in combination with TSS for the solar flare prediction models due to the high class-imbalance ratio present in the datasets.
\subsection{Evaluation}
This section presents an analysis of the results, focusing on the performance comparison of the three models used in this study. The findings reveal that the AlexNet-based model exhibits better performance in relation to both the VGG16- and ResNet34-based models, as evidenced by the HSS and TSS scores provided in Table \ref{table:scoretable}. Notably, the AlexNet-based model demonstrates enhanced robustness, as indicated by the lower standard deviation, and achieves an approximate 2\% improvement (for both TSS and HSS) compared to the VGG16-based model. Furthermore, when compared to the ResNet34-based model, the AlexNet-based model showcases a 1\% higher skill score (for both TSS and HSS). It is important to highlight that the skill scores of the VGG16 and ResNet34 models exhibit greater variability, primarily influenced by the outcomes of Fold-3 in the 4-fold cross-validation experiment, details presented in Table.~\ref{table:fold}. Furthermore, our best results exceed those reported in \cite{pandeyds2023, pandeyecml2023} by approximately 2\% in terms of TSS (reported $\sim$0.51), and in \cite{pandeyecml2023} by about 2\% in terms of HSS ($\sim$0.35), while remaining comparable to the HSS in \cite{pandeyds2023}. For reproducibility, source code for models along with results is available in our open-source project repository \cite{sourcecode}.

\begin{table}[tbh!]
\setlength{\tabcolsep}{20pt}
\renewcommand{\arraystretch}{1.5}
\vspace{-10pt}
\caption{The aggregated results in terms of TSS and HSS from test sets of a 4-fold cross-validation experiment on our models.}
\begin{center}
\begin{tabular}{c c c}
\hline
Models  & TSS & HSS\\
\hline
AlexNet & \textbf{0.526$\pm$0.05} & \textbf{0.372$\pm$0.05}\\

VGG16 & 0.506$\pm$0.09 & 0.353$\pm$0.09\\

ResNet34 & 0.513$\pm$0.09  & 0.360$\pm$0.09\\
\hline
\end{tabular}
\label{table:scoretable}
\end{center}
\vspace{-10pt}
\end{table}

\begin{table}[tbh!]
\setlength{\tabcolsep}{4pt}
\renewcommand{\arraystretch}{1.5}
\caption{Detailed results of 4-fold cross-validation experiments, showing all the four outcomes of confusion matrices (TP, FP, TN, FN) for all three models used in this study.}
\begin{center}
 \begin{tabular}{r r c c c c c c}
\hline

          %% <--  Changed
Models & Folds & TP  & FP  & TN  & FN   &TSS & HSS \\
\hline
%  &    &     &    & &    & \\
& Fold-1  &  1,729  & 2,225  & 10,229 & 605 & 0.5621 & 0.4385\\

AlexNet & Fold-2 &  1,075 & 2,298  & 11,557 & 537   & 0.5010 & 0.3380\\

& Fold-3 & 1,660 & 3,291 & 11,017 & 704 & 0.4722 & 0.3241\\ 
& Fold-4 & 2,209 & 3,549 & 10,483 & 481 & 0.5683 & 0.3890\\ 
\hline
& Fold-1  &  1,704  & 2,067  & 10,387 & 630 & 0.5641 & 0.4512\\

VGG16 & Fold-2 &  1,233 & 3,401  & 10,454 & 379   & 0.5194 & 0.2841\\

& Fold-3 & 1,409 & 3,089 &11,219 & 955 & 0.3801 & 0.2761\\ 
& Fold-4 & 2,125 & 3,236 &10,796 & 565 & 0.5593 & 0.3992\\ 

\hline
& Fold-1  &  1,779  & 2,145  & 10,309 & 555 & 0.5900 & 0.4621\\

ResNet34 & Fold-2 &  1,257 & 3,872  & 9,983 & 355   & 0.5003 & 0.2550\\

& Fold-3 & 1,328 & 2,299 & 12,009 & 1,036 & 0.4011 & 0.3280\\ 
& Fold-4 & 2,160 & 3,382 & 10,650 & 530 & 0.5620 & 0.3933\\ 
\hline
\end{tabular}
\end{center}
\label{table:fold}
\vspace{-12pt}
\end{table}

\begin{figure}[tbh!]
\centering
% \vspace{-15pt}
\includegraphics[width=0.95\linewidth ]{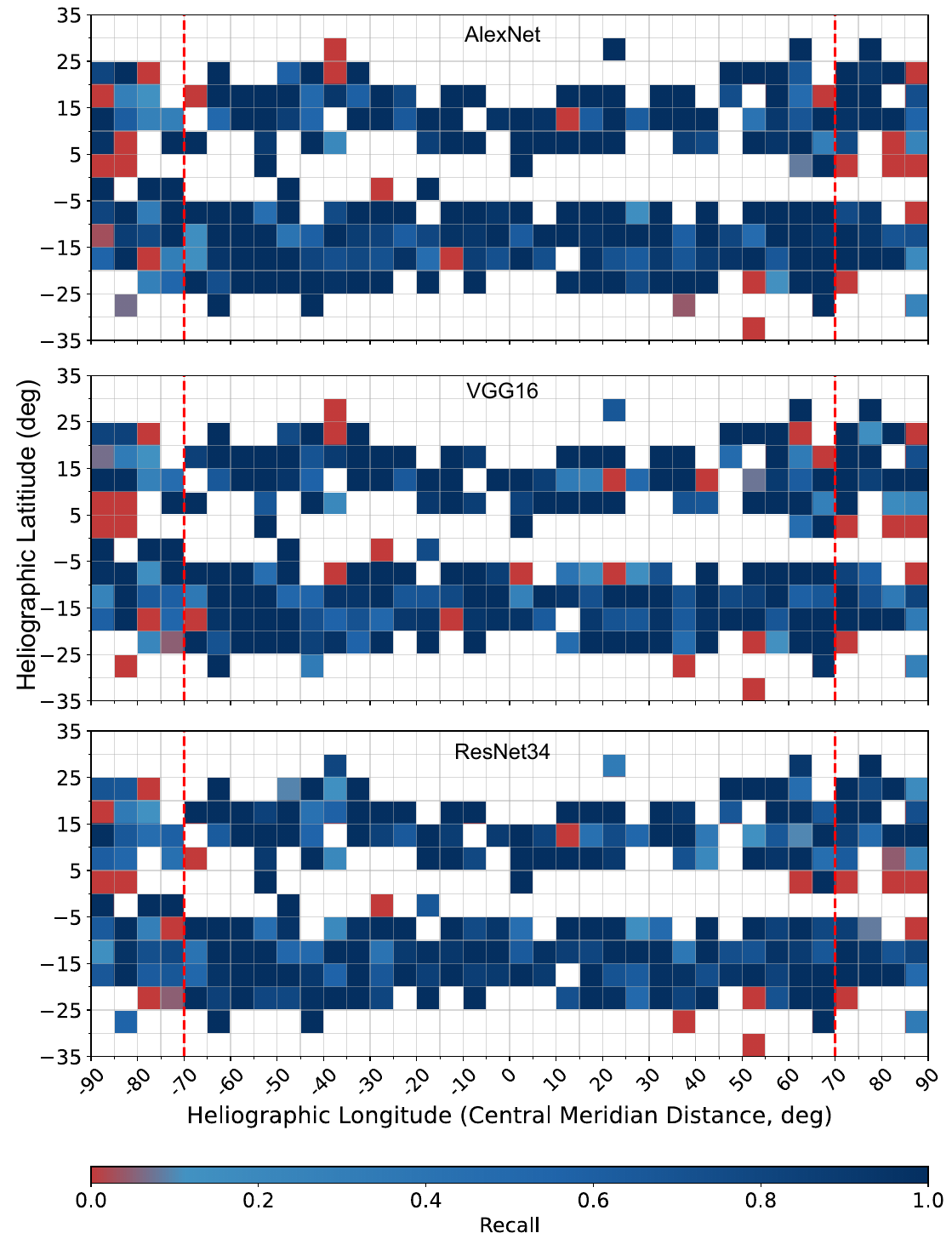}
\vspace{-10pt}
\caption[]{A heatmap illustrating all three models' recall performance for \textbf{$\geq$M-class flares} i.e., FL-class. The locations of the flares (with maximum peak x-ray flux, used as labels) are aggregated into 5$^{\circ}$ $\times$ 5$^{\circ}$ spatial bins of latitude and longitude. Note: White cells in the grid represent unavailable instances.}
\label{fig:hist_combined}
\vspace{-15pt}
\end{figure}

\begin{table}[tbh!]
\setlength{\tabcolsep}{4pt}
\renewcommand{\arraystretch}{1.5}
\caption{A comprehensive results showing correctly (TP) and incorrectly (FN) classified X- and M-class flares, distinguishing between central ($|longitude|$$\leq\pm70^{\circ}$) and near-limb locations. All counts are aggregated across the folds from the test set of the cross-validation experiment and Recall statistics are computed on the aggregated results.}
\begin{center}
 \begin{tabular}{r r c c c c c c}
\hline
 & &
\multicolumn{3}{c}{Within $\pm$70$^{\circ}$} %%\\               %% <-- mistake
%% \cline{1-2}                               %% <-- mistake
&                                            %% <--  addition
\multicolumn{3}{c}{Beyond $\pm$70$^{\circ}$}\\
          %% <--  Changed
Models & Flare-Class & TP  & FN  & Recall  & TP   &FN & Recall \\
\hline
%  &    &     &    & &    & \\
& X-Class  &  614  & 54  & \textbf{0.92} & 138 & 74 & 0.65\\

AlexNet & M-Class &  4,645 & 1,185  & \textbf{0.80} & 1,276   & 1,014 & 0.55\\

& Total (X\&M) & 5,259 & 1,239 & \textbf{0.81} & 1,414 & 1,088 & 0.57\\ 
\hline
%  &    &     &    & &    & \\
&X-Class  &  560  & 108  &  0.84 & 165 & 47 & 0.78\\

VGG16 &M-Class &  4,473 & 1,357  & 0.77 & 1,273   &1,017 & 0.55\\

&Total (X\&M) & 5,033 & 1,465 & 0.77 & 1,438 & 1,064 & 0.57\\ 

\hline
%  &    &     &    & &    & \\
&X-Class  &  612  & 56  & \textbf{0.92} & 172 & 40 & \textbf{0.81}\\

ResNet34 &M-Class &  4,449 & 1,381  & 0.76 & 1,291   &999 & \textbf{0.56}\\

&Total (X\&M) & 5,061 & 1,437 & 0.78 & 1,463 & 1,039 & \textbf{0.59}\\ 
\hline
\end{tabular}
\end{center}
\label{table:comp}
\vspace{-20pt}
\end{table}

\begin{figure*}[tbh!]
\centering
\begin{tabular}{c c}
    \includegraphics[width=0.475\linewidth]{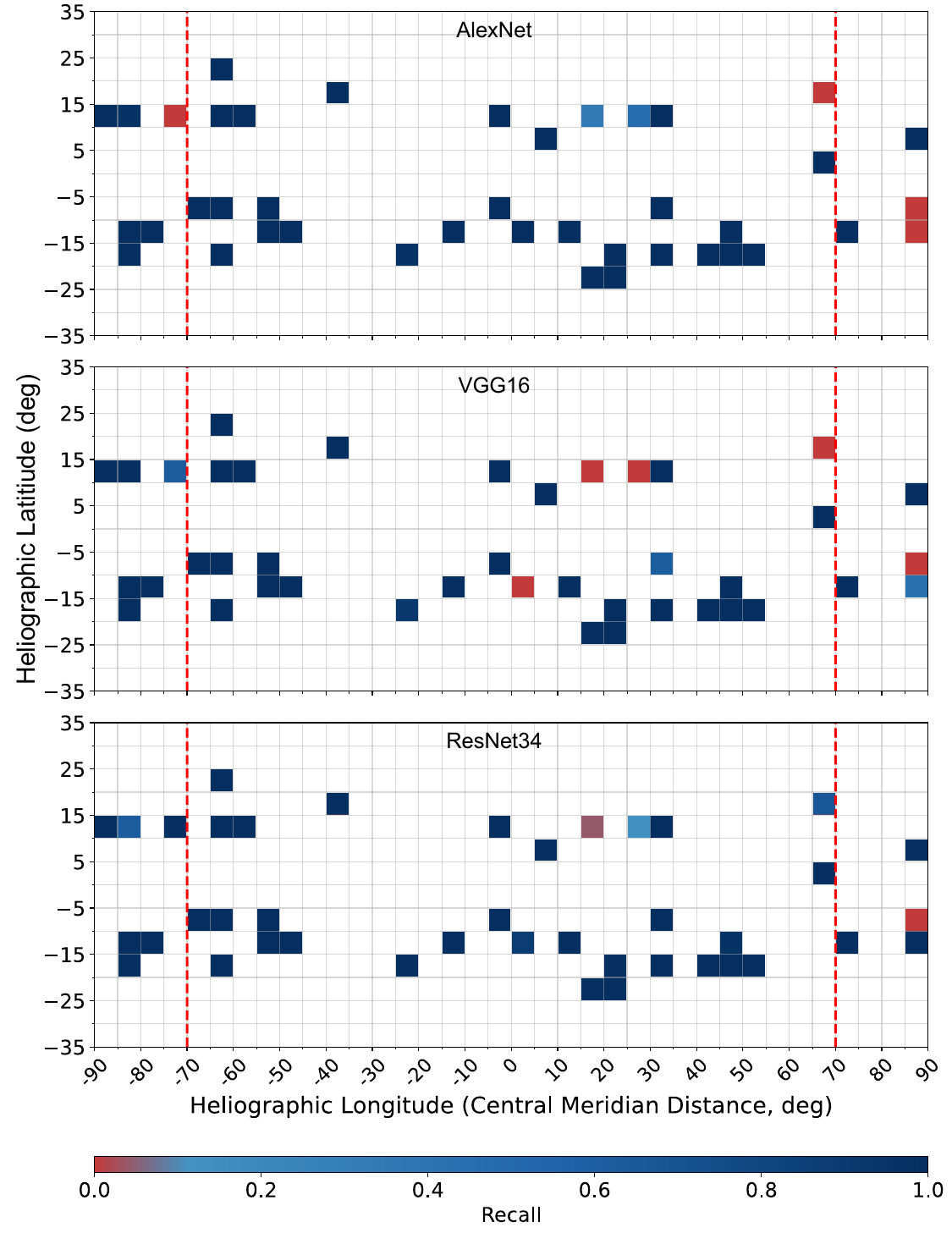} & 
\includegraphics[width=0.475\linewidth]{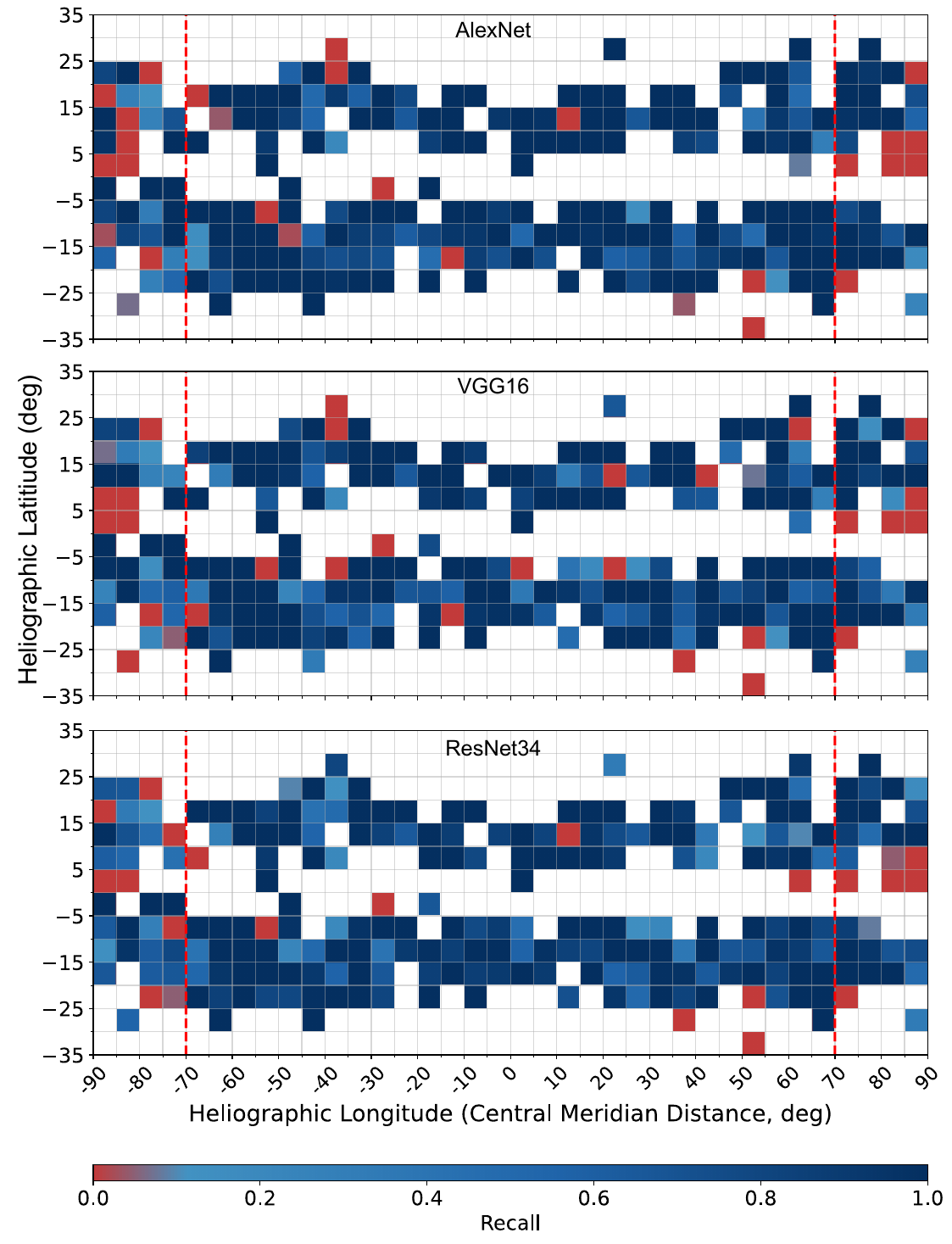} \\
    (a) & (b)
\end{tabular}

\caption{Individual heatmaps illustrating all three models' recall performance for subclasses in FL (a) \textbf{X-class flares} and (b) \textbf{M-class flares}. The spatially aggregated recall scores in Fig.\ref{fig:hist_combined} are isolated for two subclasses. White cells in the grid represent unavailable instances.}
\label{fig:histx_m}
\vspace{-15pt}
\end{figure*}

In addition, we evaluate the results by examining the correct prediction and missed flare counts for class-specific flares (X-class and M-class) in central locations and near-limb locations of the Sun, as presented in Table \ref{table:comp}. It is noteworthy that, while the overall performance measured in terms of TSS and HSS indicates the better performance of the AlexNet-based model over the other two deeper and more advanced models, VGG16 and ResNet34, the ResNet34-based model exhibits the best performance on average for near-limb events. The class-specific analysis for X- and M-class flares reveals that the ResNet34-based model achieves correct predictions for $\sim$81\% of the X-class flares ($\sim$16\% higher than AlexNet) and $\sim$56\% of the M-class flares ($\sim$1\% higher than AlexNet) in near-limb locations. Despite all models being fine-tuned with the same dataset and undergoing similar hyperparameter optimization, their distinctive architectures influenced their ability to capture specific spatial patterns and features, leading to variations in overall performance, prediction sensitivity, and recall rates for specific flare intensities and event locations. 

Moreover, we scrutinized the proficiency of our models both from a quantitative and qualitative standpoint by conducting an intricate spatial analysis of their performance in relation to the locations of M- and X-class solar flares, that were used as the labels. For the purpose of our analysis, we used the predictions made on the test set and created a heatmap by gathering the flares grouped by their location in the Heliographic Stonyhurst (HGS) coordinate system, where each bin represents a 5$^{\circ}$ $\times$ 5$^{\circ}$ spatial cell in terms of latitude and longitude. Initially, we computed the recall for the $\geq$M-class flares (combined M- and X-class flares) in each spatial cell, providing a comprehensive assessment of the models' performance. Subsequently, we evaluated the recall separately for M-class and X-class flares, allowing us to analyze the models' sensitivity at a more granular level. The heatmaps that illustrate the spatial distribution of recall scores for $\geq$M-,  X-, and M-class flares are shown in Fig.~\ref{fig:hist_combined}, ~\ref{fig:histx_m} (a), and ~\ref{fig:histx_m} (b) respectively. This allowed us to compare all three models on their capabilities to learn spatial patterns that localize the regions where the models were more effective in making accurate predictions and vice versa.

% \begin{figure}[htbp!]
% \centering
% \includegraphics[width=0.95\linewidth ]{figures/x_hgs.pdf}
% \caption[]{A heatmap illustrating the quantitative and qualitative evaluation of all three models' recall performance for \textbf{X-class flares}. The locations of the flares (with maximum peak x-ray flux, used as labels) are aggregated into 5$^{\circ}$ $\times$ 5$^{\circ}$ spatial bins of latitude and longitude. Note: Red cross in white grids represents locations with zero correct predictions while white cells without red cross represent unavailable instances.}
% \label{fig:hist_x}
% \end{figure}

% \begin{figure}[htbp!]
% \centering
% \includegraphics[width=0.95\linewidth ]{figures/m_hgs.pdf}
% \caption[]{A heatmap illustrating the quantitative and qualitative evaluation of all three models' recall performance for \textbf{M-class flares}. The locations of the flares (with maximum peak x-ray flux, used as labels) are aggregated into 5$^{\circ}$ $\times$ 5$^{\circ}$ spatial bins of latitude and longitude. Note: Red cross in white grids represents locations with zero correct predictions while white cells without red cross represent unavailable instances.}
% \label{fig:hist_m}
% \end{figure}

Our findings indicate that all three models demonstrated reasonable proficiency in predicting X-class flares in central locations. However, among these, the ResNet34-based model stood out for its overall better performance in accurately forecasting X-class flares, regardless of whether they were in near-limb or central locations as shown in Fig.~\ref{fig:histx_m} (a). Upon analysis of the heatmaps for $\geq$M- and only M-class flares, as depicted in Fig.~\ref{fig:hist_combined} and Fig.~\ref{fig:histx_m} (b) respectively, we observed that the ResNet34-based model generally yielded more accurate predictions across diverse spatial locations in comparison to the other models. Nonetheless, a common limitation across all three models was an elevated rate of false negatives in near-limb areas for M-class flares. Notably, these regions are often associated with unreliable readings due to projection effects. Despite this challenge, our study signifies a substantial progression in solar flare prediction, enabling the prediction of flares even in these intricate near-limb regions with distorted magnetic fields. This ability to predict flares in near-limb areas has considerable implications in operational forecasting. 
\section{Conclusion and Future Work} \label{sec:conc}
Our study involved the evaluation of three deep learning models, namely AlexNet, VGG16, and ResNet34, for the prediction of solar flares, with a primary focus on their performance for near-limb events. Through rigorous evaluation of our models and analysis of the prediction results, we observed a notable performance advantage of the ResNet34-based model in predicting near-limb flares. This finding highlights the efficacy of employing deeper architectures with residual connections, which enhance feature extraction and facilitate the capture of subtle patterns associated with near-limb events. Moreover, our study highlighted the variability in model performance across different flare types and event locations, emphasizing the importance of tailoring models and analyzing results in context-specific manners. This shows the need for further exploration of deep learning models to effectively capture the diverse nature of flare events. The implications of our research extend to operational forecasting systems, where the precise and reliable prediction of solar flares, including near-limb events, holds significant importance. Future research directions can explore the integration of multi-modal data and the development and utilization of deep learning models that can learn from temporally evolving solar activity. \\

\noindent \textbf{Acknowledgements:} This work is supported in part under two NSF grants (Award \#2104004 and \#1931555) and a NASA SWR2O2R grant (Award \#80NSSC22K0272).
% Data used in this study is a courtesy of NASA/SDO and the AIA, EVE, and HMI science teams, and the NOAA (NGDC).
% , the interpretability of learned features, and the utilization of explainable deep learning techniques to enhance predictive capabilities and address limitations. 
% The improved capabilities demonstrated by our models provide valuable insights for refining forecasting methodologies in operation and facilitating real-time decision-making processes.
% Apart from the promising capabilities of our models, it is important to highlight the associated inherent challenges. Factors such as data availability, observational constraints, and the evolving nature of solar activity pose ongoing obstacles to model development and validation. Addressing these challenges necessitates advancements in data collection, integration, and the development of sophisticated models. Future research directions can explore the integration of multi-modal data, the development of models that can capture temporally evolving solar activity, the interpretability of learned features, and the utilization of explainable deep learning techniques to enhance predictive capabilities and address limitations. 
% Overall, our study contributes to the growing body of research in solar flare prediction, shedding light on the capabilities and limitations of different model architectures, particularly for near-limb flares. These insights hold the potential to drive advancements in ultimate operational forecasting efforts. 
\bibliographystyle{IEEEtran}
\bibliography{references}
\end{document}